\documentclass[pdfusetitle,aps,prl,nofootinbib,twocolumn,superscriptaddress,preprintnumbers]{revtex4-2}
\pdfoutput=1
\usepackage{amssymb}
\usepackage{amsmath}
\usepackage[svgnames,dvipsnames,x11names]{xcolor}
\usepackage{hyperref}
\hypersetup{
colorlinks=true,
citecolor= BrickRed,
linkcolor= OliveGreen,
urlcolor= BrickRed,
pdfauthor={},
pdftitle={},
pdfsubject={}
}

\def \eps {{\epsilon}}
\def \veps {{\vec \eps}}
\def \k {{\vec k}}
\def \K {{\vec K}}
\def \z {{\vec z}}

\def \hF {{\widehat F}}
\def \hS {{\widehat S}}
\def \hW {{\widehat W}}
\def \hD {{\widehat D}}
\def \hX {{\widehat X}}
\def \hn {{\widehat n}}

\def \O {{\cal O}}

\def \LA {{\langle}}
\def \RA {{\rangle}}

\def\nn{\nonumber\\}

\begin{document}

\title{Cosmological Double-Copy Relations}

\author{Hayden Lee}
\affiliation{Kavli Institute for Cosmological Physics, University of Chicago, Chicago, IL 60637, USA}
\author{Xinkang Wang}
\affiliation{Department of Physics, University of Chicago, Chicago, IL 60637, USA}

\begin{abstract}
	We present differential double-copy relations between gluon and graviton three-point functions in (A)dS$_{d+1}$. 
	We introduce a set of differential operators in (A)dS that naturally generalize on-shell kinematics of scattering amplitudes in flat space.
	This provides a way to construct (A)dS correlators by replacing the kinematic variables of amplitudes with the corresponding differential operators and suitably ordering them. 
	By construction, the resulting correlators are manifestly conformally invariant, with the correct flat-space limit, and exhibit a differential double-copy structure.
\end{abstract}

\maketitle

\section{Introduction}

Correlation functions in an approximate de Sitter (dS) space are the fundamental observables of inflationary cosmology. 
At the heart of standard Lagrangian calculations of inflationary correlators are time integrals, which track local time evolution of quantum fields in the bulk spacetime. 
While evaluating these time integrals is notoriously complicated, the resulting spatial correlators greatly simplify when taken to the future boundary of dS. 
It is then natural to wonder whether there is a radically different way of computing these boundary correlators, without any reference to bulk time evolution.

The past several years have seen an intensive focus on the study of cosmological correlators from a boundary perspective.
In this framework, basic physical principles such as symmetry and unitarity are used as fundamental inputs to determine the final observables, rather than arising as nontrivial outputs of a calculation~\cite{Maldacena:2011nz,Raju:2012zr, Raju:2012zs, Mata:2012bx, Bzowski:2013sza, Kundu:2014gxa, Arkani-Hamed:2015bza, Kundu:2015xta, Arkani-Hamed:2017fdk, Arkani-Hamed:2018kmz, Arkani-Hamed:2018bjr, Sleight:2019mgd, Sleight:2019hfp, Baumann:2019oyu, Green:2020ebl, Baumann:2020dch, Goodhew:2020hob, Pajer:2020wxk, Jazayeri:2021fvk, Melville:2021lst, Bonifacio:2021azc, Sleight:2021plv, Baumann:2021fxj, Hogervorst:2021uvp, DiPietro:2021sjt, Sleight:2021iix, Bzowski:2022rlz}. 
The ongoing program of the {\it cosmological bootstrap} (see~\cite{Benincasa:2022gtd,Baumann:2022jpr} for reviews) has revealed the underlying analytic structure of cosmological correlators, and powerful new ways of computing them, that are highly obscure from the Lagrangian formalism.

Both the philosophy and technology of the cosmological bootstrap are heavily inspired by the modern on-shell program of scattering amplitudes~\cite{Elvang:2013cua,Dixon:2013uaa, Cheung:2017pzi, Travaglini:2022uwo}.
The fact that many useful tools for scattering amplitude calculations can be applied to the cosmological context is not a mere coincidence. 
In momentum space, a direct connection between cosmological correlators in dS and scattering amplitudes in flat space is furnished by the {\it total energy singularity}~\cite{Maldacena:2011nz, Raju:2012zr}, which is the analog of a bulk-point singularity in Lorentzian anti-de Sitter (AdS) space~\cite{Gary:2009ae, Heemskerk:2009pn, Maldacena:2015iua}. Essentially, cosmological correlators arising from local bulk dynamics must reduce to amplitudes in the limit when the sum of external energies goes to zero in the complex energy plane~\cite{Arkani-Hamed:2015bza, Arkani-Hamed:2017fdk, Arkani-Hamed:2018kmz, Arkani-Hamed:2018bjr}, which allows us to think of cosmological correlators as a particular deformation of scattering amplitudes away from the singular locus. 
This raises a tantalizing prospect that many of the remarkable properties of scattering amplitudes can be generalized to cosmological correlators. 

One of the most striking features of amplitudes is the double-copy relation between gauge and gravity theories, which expresses graviton amplitudes as two copies of gluon amplitudes. 
After its original discovery in string theory~\cite{Kawai:1985xq}, this relation has been extended to amplitudes at higher multiplicities and multiple loops~\cite{Bern:2008qj,Bern:2010ue}, to scalar and supersymmetric theories~\cite{Bern:2011rj, Cachazo:2014xea}, and has also found applications in gravitational-wave physics (see \cite{Bern:2019prr,Bern:2022wqg, Adamo:2022dcm} for reviews). 
A natural question is then whether there exists a generalized notion of double copy in curved backgrounds. It remains technically challenging to compute graviton correlators in (A)dS beyond three points~\cite{Raju:2012zs, Bonifacio:2022vwa}, and therefore an extension of double copy beyond flat space will be highly valuable.

An important lesson from the Bern-Carrasco-Johansson (BCJ) construction of the double copy~\cite{Bern:2008qj,Bern:2010ue} is that the right objects to be double copied are the special combinations of kinematic variables that obey the Jacobi relation.
This motivates a similar strategy in (A)dS. That is, to first identify the right kinematic building blocks for correlators. 
In~\cite{Arkani-Hamed:2018kmz, Baumann:2019oyu, Baumann:2020dch}, so-called {\it weight-shifting operators}---differential operators that shift quantum numbers in conformal field theories~\cite{Costa:2011dw, Karateev:2017jgd}---were developed in the context of cosmology.
This approach highlighted the fact that differential operators can be used as basic building blocks to generate spinning correlators from simpler scalar correlators. 
A similar approach was used in~\cite{Roehrig:2020kck,Eberhardt:2020ewh, Herderschee:2022ntr, Diwakar:2021juk, Sivaramakrishnan:2021srm,  Cheung:2022pdk}, showing that exchange diagrams in AdS can be expressed as differential operators acting on a scalar contact diagram. 
In particular, these recent developments have uncovered the curved-space generalization of the double copy of scalar theories.
Yet, a double-copy formulation of spinning correlators has so far remained elusive, even at the three-point level  (see~\cite{Li:2018wkt,Farrow:2018yni,Lipstein:2019mpu,Armstrong:2020woi,Albayrak:2020fyp,Alday:2021odx,Diwakar:2021juk,Zhou:2021gnu, Sivaramakrishnan:2021srm,Jain:2021qcl, Cheung:2022pdk,Herderschee:2022ntr,Armstrong:2022csc} for recent investigations).

In this letter, we present new differential representations of the gluon and graviton three-point functions in $d$-dimensional (A)dS space.\footnote{Specifically, we consider Euclidean AdS correlators and dS wavefunction coefficients on the respective boundaries, which have the same kinematic structure up to overall normalization factors that we drop.}
We first determine conformally-invariant differential operators that serve as kinematic building blocks for spinning conformal correlators. 
We find that these operators, when suitably ordered, become natural generalizations of the kinematic variables of spinning amplitudes to (A)dS space. 
This mapping between the basic kinematic structures allows us to promote flat-space spinning amplitudes to the corresponding (A)dS correlators in a straightforward fashion. 
We construct the three-point functions of gauge and gravity theories in this way, and show that their kinematic building blocks exhibit a manifest double-copy structure.

\section{Correlator Building Blocks}

We focus our attention to correlators of conserved currents on the boundary, which are dual to massless spinning particles in the bulk. 
Two important physical criteria for these correlators are conformal invariance and current conservation, which are the analogs of Lorentz invariance and on-shell gauge invariance for amplitudes.  
To solve the symmetry constraint, we will use the weight-shifting operators developed in~\cite{Costa:2011dw, Karateev:2017jgd,Costa:2018mcg,Baumann:2019oyu}, which are conformally-covariant differential operators that transform in finite-dimensional representations of the conformal algebra.
These operators are naturally constructed using the embedding space formalism~\cite{Dirac:1936fq,Costa:2011mg}, where conformal transformations in $\mathbb{R}^{d}$ are realized as Lorentz transformations on a higher-dimensional lightcone embedded in $\mathbb{R}^{1, d+1}$.

To make a direct connection with scattering amplitudes, we consider the momentum-space version of the weight-shifting operators. 
For three-point functions, we find it most useful to consider the following set of operators~\cite{Baumann:2019oyu}:\footnote{In~\cite{Costa:2011dw, Baumann:2019oyu}, the notation $D_{aa}$ was used for $F_{ab}$ with a fixed $b$, e.g.,~$D_{11}=F_{12}$ and $D_{22}=F_{21}$. Here, we will allow for different choices of $b$.}
\begin{align}
	S_{ab}&\equiv\rho_a\rho_b(\z_a\cdot \z_b) +(\z_b\cdot \k_b)D_{a b}+(\z_a\cdot \k_a)D_{b a}\nn
	&\quad +(\z_a\cdot \k_a)(\z_b\cdot \k_b)W_{ab}\,, \label{Sab} \\
	D_{a b} &\equiv \rho_a(\z_a\cdot \K_{ab}) - (\z_a\cdot \k_a)W_{ab}\, ,\\
	F_{ab} &\equiv (\k_b\cdot \K_{ab}+\Delta_b-d)\z_a\cdot \K_{ab}-(\z_b\cdot \K_{ab})(\z_a\cdot\partial_{\z_b})\nn
	& \quad+(\z_a\cdot \z_b) \partial_{\z_b}\cdot\K_{ab} -(\z_a\cdot \k_b)W_{ab}\, ,\label{Fab}\\
	W_{ab} &\equiv\frac{1}{2}\K_{ab}\cdot \K_{ab}\, ,\quad \K_{ab} \equiv \partial_{\k_a}-\partial_{\k_b}\, ,
\end{align}
where $\k_a$ is the momentum, $\z_a$ is an auxiliary null vector, and ${\rho_a\equiv \Delta_a+\ell_a-1}$, with $\Delta_a$ denoting the weight and $\ell_a$ the spin of a conformal primary $\O_{\Delta_a,\ell_a}$. The subscripts $a,b=1,2,3$ label the position, the arrow over a variable denotes a vector in $\mathbb{R}^d$, and a dot product indicates the Euclidean inner product of two vectors with the metric $\delta_{\mu\nu}$, with $\mu,\nu$ labelling spatial indices in $\mathbb{R}^d$.
The operators above have the following action: the {\it spin operator} $S_{ab}$ raises the spin at points $a$ and $b$ by one unit, the {\it weight operator} $W_{ab}$ lowers the weights at points $a$ and $b$ by one unit, and the {\it spin-weight operators} $D_{a b}$ and $F_{ab}$ both raise the spin at point $a$ by one unit, while lowering the weight at point $b$ and $a$ by one unit, respectively. 
The corresponding embedding-space expressions of these operators can be found in~\cite{Karateev:2017jgd, Baumann:2019oyu}. 

We are naturally led to the question: What is the flat-space limit of the weight-shifting operators? 
As we will see, this knowledge will enable us to directly construct (A)dS correlators given the corresponding amplitudes in flat space via appropriate replacements of kinematic building blocks.
It turns out that the operators above are direct analogs of polarization vector and momentum contractions in flat space such as $\veps_a\cdot\veps_b$ and $\veps_a\cdot \k_b$. Seeing this requires a proper normalization and ordering of these operators, which we discuss next.

\section{Amplitude-Correlator Dictionary}

In the weight-shifting approach, boundary spinning three-point functions in (A)dS$_{d+1}$ are represented as
\begin{equation}
	\LA J_\ell J_\ell J_\ell\RA = \hn_\ell\LA\Phi\Phi\Phi\RA\, , \label{JlJlJl}
\end{equation}
where $\hn_\ell$ represents a combination of weight-shifting operators, $J_\ell$ is a spin-$\ell$ conserved tensor with $\Delta_{J_\ell} = d+\ell-2$, and $\Phi$ is an integer-weight scalar dual to a shift-symmetric bulk scalar $\phi$~\cite{Bonifacio:2018zex,Blauvelt:2022wwa}. We use the index-free notation $J_\ell \equiv \eps_{\mu_1}\cdots \eps_{\mu_\ell}J_\ell^{\mu_1\cdots\mu_\ell}$ with the indices contracted with the polarization vector $\veps$. 
The weight-shifting operators are non-singular, whereas the three-point function of $\Phi$ diverges as $K^{\frac{3-d}{2}}$ for $d>3$ (or $-\log K$ for $d=3$):
\begin{equation}
	\lim_{K\to 0}\LA\Phi\Phi\Phi\RA  = A_{\phi^3}\times (k_1k_2k_3)^{\Delta_\Phi-\frac{d+1}{2}}K^{\frac{3-d}{2}}\, ,\label{Phiflat}
\end{equation}
where $k_a \equiv |\k_a|$ denotes the energy at point $a$, ${K\equiv k_1+k_2+k_3}$ is the total energy, and we have suppressed the delta function that enforces spatial momentum conservation.
The coefficient $A_{\phi^3}$ is the corresponding amplitude in flat space, which in this case is just a constant that we will set to unity, $A_{\phi^3}=1$. 

Due to the inherent non-commutativity of weight-shifting operators, the differential representation \eqref{JlJlJl} is far from unique. 
A widely-used strategy is to enumerate all possible combinations of operators and fix their coefficients by other dynamical constraints such as imposing the correct behavior in the flat-space limit. 
However, naively applying this procedure generically leads to representations of correlators that are both algebraically cumbersome and physically unintuitive, obscuring their connection to scattering amplitudes.

In fact, there is a canonical normalization and ordering of operators that most directly reveals the flat-space limit.  
First of all, it turns out that it is most natural to have all the weight operators to act on the scalar correlator first. 
This is due to the special property of $W_{ab}$ that it does not change the degree of singularity in $K$ when acting on a function that goes as $K^{\frac{3-d}{2}}$, which is precisely the behavior of the scalar seed function in \eqref{Phiflat}. In other words,
\begin{equation}
	\lim_{K\to 0}W_{ab} (f K^{\frac{3-d}{2}}) =  \lim_{K\to 0}(W_{ab} f)K^{\frac{3-d}{2}}\, ,
\end{equation}
with $f$ some function of momenta. Let us define the normalized version of the operator as
\begin{equation}
	\hW_{ab} \equiv -\frac{2W_{ab}}{(\Delta_a+\Delta_b-\Delta_c - 2)(\Delta_a+\Delta_b-\tilde\Delta_c - 2)}\,,\label{Wnorm}
\end{equation}
with $c\ne a,b$, where $\tilde\Delta_c=d-\Delta_c$ the shadow weight. 
This choice ensures unit normalization in the flat-space limit and takes into account the weights for which the singularity in $K$ vanishes after acting with $W_{ab}$. 
Another advantage of acting first with $W_{ab}$ is that this avoids acting on the longitudinal factors $\z_a\cdot \k_a$ in the other weight-shifting operators. 
These factors vanish when we evaluate the correlator {\it on-shell}, by which we mean computing the transverse-traceless part of the correlator with $\z_a$ replaced by the physical polarization vectors $\veps_a$. 

Next, consider the spin operator $S_{ab}$. This has a non-derivative term that becomes $\veps_a\cdot\veps_b$ on-shell, while all of its derivative terms get multiplied by longitudinal factors. Consequently, the spin operators have a very simple on-shell action
\begin{equation}
	\hS_{a_1b_1}\cdots \hS_{a_nb_n}|_{z\to \eps}   = (\veps_{a_1}\cdot \veps_{b_1})\cdots(\veps_{a_n}\cdot \veps_{b_n}) \, ,
\end{equation}
when no other operators act on them, where we have normalized the operator as
\begin{equation}
	\hS_{ab} \equiv \frac{1}{\rho_a\rho_b}S_{ab}\, .
\end{equation}
It is thus most natural to act with the spin operators last, in which case they simply turn into a product of polarization factors.

It remains to discuss the spin-weight operators $D_{a b}$ and $F_{a b}$. While their on-shell actions are less trivial, it turns out that they both turn into $\veps_a\cdot \k_b$ in the flat-space limit (rescaled by energy factors). To see this, consider the on-shell action of two $D_{a b}$ operators, which can be expressed in terms of energy derivatives as
\begin{align}
	\hD_{a b}\hD_{c d}|_{z\to\eps} 
	& =\frac{(\veps_{a}\cdot  \k_b)(\veps_{c}\cdot  \k_d)}{k_bk_d}\left(\partial_{ k_b} \partial_{k_d}-\frac{\delta_{bd}}{k_d}\partial_{k_d}\right) \nn
	 &\hskip -40pt +(\veps_{a}\cdot\veps_{c})\left[\frac{\delta_{bd}-\delta_{ad}}{k_d}\partial_{k_d}-\delta_{bc}\left(\frac{\partial_{k_c}}{k_c}+\frac{W_{cd}}{\rho_{c}}\right)\right] ,\label{DD}
\end{align}
where $\delta$ is the Kronecker delta and we have normalized the operator as 
\begin{equation}
	\hD_{ab} \equiv \frac{1}{\rho_a} D_{ab}\, .
\end{equation}
The two-derivative term in the first line of \eqref{DD} gives the most singular term in $K$, and reduces to the aforementioned kinematic structure in the flat-space limit. The other terms in \eqref{DD} have different consequences depending on the index permutations of the operators.
To see why, consider correlators of conserved currents in odd~$d$. These are rational functions of energies, whereas the scalar seeds always have a logarithmic singularity.
The spin-weight operators must then combine to remove this logarithmic singularity, which implies a set of selection rules for index permutations that can appear. 
For instance, the one-derivative term in the first line of \eqref{DD} gives a logarithmic singularity that cannot be canceled against other terms due to its polarization structure, which forbids the operator combinations such as $\hD_{13}\hD_{23}$, for which $b=d$. 
Similarly, $F_{ab}$ has the same kinematic structure as $D_{ab}$ in the flat-space limit due to the fact that $\k_b\cdot \K_{ab}$ in \eqref{Fab} does not increase the degree of singularity in $K$. Its normalized version is given by
\begin{align}
	\hF_{ab} \equiv \frac{1}{\Delta_a+\ell_b+\ell_c-2}F_{ab}\, .
\end{align}
Similar to \eqref{Wnorm}, this takes into account the spin and weight combinations for which correlators become trivial.

We will refer to the ordering $\hS\cdots \hS\hX\cdots \hX\hW\cdots \hW$ of the weight-shifting operators as {\it normal ordering}, where $\hX\in\{\hD,\hF\}$.
As we discussed, this has the convenient feature that the operators essentially become multiplicative when evaluated on-shell and makes it easy to track the singularity structure. 
These properly normalized, normal-ordered, weight-shifting operators then serve a dual purpose: they trivialize both conformal symmetry and the flat-space limit.
In particular, we have the following dictionary between the kinematic variables for amplitudes and the normalized weight-shifting operators:\footnote{Ref.~\cite{Cheung:2017ems} studied differential operators for amplitudes that strip off contractions of polarization vectors and momenta. The dictionary here implies that their (A)dS analogs are functional derivatives with respect to the weight-shifting operators.}
\begin{equation}
	\veps_a\cdot\veps_b  \leftrightarrow  \hS_{ab}\, ,\quad 
	  \veps_a\cdot \k_b \leftrightarrow  \hD_{ab}\hskip 1pt, \hF_{a b}\, , \quad 1  \leftrightarrow  \hW_{ab} \, ,\label{map}
\end{equation}
when normal-ordered. While the weight operators reduce to unity in the flat-space limit, they need to be suitably inserted in correlators to give the correct scaling weights. As we describe below, the choice between $\hD_{ a b}$ and $\hF_{ a b}$ depends on the type of interactions under consideration. Note that for $\veps_a\cdot \k_b$, this is in fact a one-to-two mapping; $\veps_1\cdot \k_2=-\veps_1\cdot \k_3$ but $\hD_{12}\ne -\hD_{13}$ away from $K=0$. However, these two permutations typically only differ by a local term that has a delta-function support in position space, which is the boundary manifestation of the field redefinition freedom in the bulk. Flat-space amplitudes are of course invariant under any field redefinitions, and local terms do not survive in the flat-space limit because they do not have any singularities.

\section{Three-Point Double Copy}

The three-particle amplitudes for Yang-Mills (YM) theory and general relativity (GR) take the form\footnote{To make a direct comparison with correlators in the flat-space limit, we have shown the amplitudes computed in axial gauge, and also suppressed the coupling constants and color factors.}
\begin{align}
	A_{\rm YM} &=  (\veps_1\cdot  \veps_2)(\veps_3\cdot \k_1)+\text{cyc.}\, ,\quad\hskip 5pt A_{\rm GR} = A_{\rm YM}^2\,, \label{AYM}\\
	A_{F^3} &=  (\veps_1\cdot \k_2)  (\veps_2\cdot \k_3)(\veps_3\cdot \k_1) \,, \quad	A_{W^3} = A_{F^3}^2 \,, \label{AF3}
\end{align}
where the first line shows the pure YM and GR amplitudes, while the second line shows the amplitudes from the higher-derivative interactions $F^3$ and $W^3$, with $F$ the YM field-strength tensor and $W$ the Weyl tensor. 
We see that the three-point amplitudes exhibit manifest double-copy relations between gauge and gravity theories. 
In this section, we present similar differential double-copy relations for spinning three-point functions in (A)dS space.

\subsection{YM and GR}

Let us first consider the three-point function of conserved spin-1 currents dual to bulk gluons. The idea is to promote the amplitude building blocks in \eqref{AYM} to differential operators via the dictionary~\eqref{map}. 
This turns $(\veps_1\cdot  \veps_2)(\veps_3\cdot \k_1)$ into, e.g.,~the spin-raising combination $\hS_{12}\hD_{3 1}$, which lowers the weight at point 1 by one unit.
To land on the correct weight for the conserved spin-1 current $\Delta_{J_1} = d-1$ at all three points, a natural seed object to use is the massless scalar three-point function $\LA\Phi\Phi\Phi\RA$ with $\Delta_\Phi=d$ accompanied by $\hW_{23}$.
This allows us to write\footnote{In $d=3$, there is a somewhat simpler momentum-space representation given by
\begin{equation}
 \LA J_1J_1J_1\RA|_{d=3}  = (k_1\hS_{12}\hD_{31}+\text{cyc.})\LA \varphi\varphi\varphi\RA\, ,
\end{equation}
where $\varphi$ is dual to a conformally coupled scalar with $\Delta_\varphi=\Delta_{J_1}=2$, which implies that the weight-shifting combination has an overall weight of zero. 
This representation is not suitable for double copy, however, since the multiplication by $k_a$ is the shadow transform of $J_1$, which becomes an integral in position space, and $k_a^2$ is not the shadow transform of $J_2$. 
}
\begin{equation}
 \LA J_1J_1J_1\RA  = \underbrace{(\hS_{12}\hD_{31}\hW_{23}+\text{cyc.})}_{\equiv\,\hn_1}\LA \Phi\Phi\Phi\RA_{\Delta_\Phi=d}\, .\label{JJJ1}
\end{equation}
By construction, this is conformally invariant and has the correct flat-space limit. 
One still needs to check the current conservation condition, which requires the correlator to be annihilated by the divergence operator in embedding space~\cite{Costa:2011mg}
\begin{align}
	\text{div}_{a} &\equiv \partial_{X_a}\cdot T_{Z_a}\, ,\\
	T_{Z_a} & \equiv\left(\frac{d}{2}-1+Z_a\cdot\partial_{Z_a}\right)\partial_{Z_a}-\frac{1}{2}Z_a \partial_{Z_a}\cdot \partial_{Z_a}\, ,
\end{align}
where $X_a$ is an embedding-space coordinate and $Z_a$ is an auxiliary null vector in $\mathbb{R}^{1,d+1}$, which are related to $\vec x_a$ (position-space coordinate conjugate to $\k_a$) and $\z_a$ upon projection to $\mathbb{R}^d$.
The equivalent condition in momentum space is the Ward-Takahashi (WT) identity~\cite{Bzowski:2013sza,Bzowski:2017poo, Baumann:2020dch,Baumann:2021fxj}, which relates the longitudinal part of a correlator to lower-point functions. It can be checked that \eqref{JJJ1} is indeed divergenceless in general dimensions.

We now come to our double-copy construction of the graviton three-point function. 
Note that the naive procedure of squaring the whole correlator would not work for the following reasons. 
First, since $\Delta_{J_2}=\Delta_{J_1}+1$ and the operator $\hn_1$ has an overall scaling weight of $-1$, we need to accordingly adjust the weight of the seed scalar from $\Delta_\Phi=d$ to $\Delta_\Phi=d+2$. 
Another important subtlety is that conformal symmetry combined with the flat-space limit does not fully guarantee that the resulting correlator satisfies the WT identity.
As we described before, only certain operator combinations cancel the undesired singularity of the scalar seed.
To see this, note that $\hS_{12}\hS_{23}\hS_{31}\LA\Phi\Phi\Phi\RA|_{\Delta_\Phi=d}$ is conformally invariant and has the correct quantum numbers of a conserved spin-2 three-point function, and so it can in principle be part of the correlator. However, it has an unphysical, lower-order singularity, which is not constrained by the flat-space limit.

Taking these considerations into account, we have found that the graviton three-point function admits the following representation:
\begin{equation}
 \LA J_2J_2J_2\RA  =\, \underbrace{:\hn_1^2:}_{\equiv\, \hn_2}\LA \Phi\Phi\Phi\RA_{\Delta_\Phi=d+2}\, ,\label{JJJ2}
\end{equation}
where operators enclosed within colons are normal-ordered, with $\hD_{a b}\hD_{c d}$ ordered such that ${a\le c}$. This ordering of the operators ensures the cancellation of the undesired singularity of the scalar seed. Explicitly, we have
\begin{align}
	&\hn_2|_{\hW=1} = \hS_{12}^2\hD_{3 1}^2+\hS_{23}^2\hD_{1 2}^2+\hS_{31}^2\hD_{2 3}^2 \label{n2} \\[2pt]
&+2(\hS_{12}\hS_{23}\hD_{1 2}\hD_{3 1}+\hS_{12}\hS_{31}\hD_{2 3}\hD_{3 1}+\hS_{23}\hS_{31}\hD_{1 2}\hD_{2 3} )\, .\nonumber
\end{align}
To avoid clutter we have only shown part of the formula after stripping off various factors of the weight operators on the right. To reintroduce them, note that each $\hD_{ab}$ in \eqref{JJJ1} is accompanied by $\hW_{cd}$ with ${b\ne c\ne d}$.
We see that the kinematic operator $\hn_2={:\hn_1^2:}$ exhibits a double-copy structure, akin to the amplitude \eqref{AYM}. Namely, the expression \eqref{n2} can be recognized as the square of a multinomial, cf.~${(r+s+t)^2=r^2+s^2+t^2+2(rs+rt+st)}$. 
In the flat-space limit, the correlator directly reduces to the amplitude $A_{\rm GR}$, as implied by the dictionary~\eqref{map}. 
Again, it can be checked that \eqref{JJJ2} is divergenceless in general dimensions.

The differential representation is not unique, even when the operators are normal-ordered. This is due to the non-vanishing commutator ${[\hD_{a b},\hD_{c d}]\ne 0}$ for $a\ne c$. (In contrast, $\hS_{ab}$ and $\hW_{ab}$ have vanishing commutators among themselves.) 
We may also take two copies of $\hn_1$ with different permutations, which gives the same non-local part of the correlator, but can differ by local terms in momentum space. 
For instance, there exists a cyclic-symmetric representation of operators given by
\begin{align}
	&\hn_2^{\rm cyc.}|_{\hW=1} \equiv \hS_{12}^2\hD_{3 1}^2 \nn
	&\quad + 2\hS_{12}\hS_{13} \frac{{\hD_{2 3}\hD_{3 1}\!+\!\hD_{2 1}\hD_{3 2}\!-\!\hD_{2 3}\hD_{3 2}}}{3} +\text{cyc.}\, ,\label{D2b}
\end{align}
which differs from \eqref{n2} by a local term after acting on the scalar seed.
In $d=3$ momentum space, this representation precisely reproduces the graviton three-point function computed in~\cite{Maldacena:2002vr, Maldacena:2011nz}. For concreteness, let us also provide the expressions for $d=5,7$ obtained from \eqref{D2b}:\footnote{\label{foot6}The flat-space limit of spinning three-point functions that is consistent with our normalization convention is
\begin{align}
	\lim_{K\to 0}\LA J_{\ell} J_{\ell} J_{\ell}\RA = A_{\ell}\times\frac{(k_1k_2k_3)^{\frac{d+2\ell-5}{2}}}{(\delta_{d,3}+\frac{d-3}{2})(\frac{d-1}{2})_{\ell-1}}K^{\frac{3-d}{2}-\ell}\, ,\label{JJJflat}
\end{align}
where $(a)_n$ is the Pochhammer symbol and $A_{\ell}$ is the corresponding spin-$\ell$ amplitude in flat space. For three-point functions from higher-derivative interactions~\eqref{hd}, the scaling instead becomes $K^{\frac{3-d}{2}-3\ell}$. For even $d$, the correlators of conserved currents contain branch cuts, and taking the flat-space limit requires a more detailed analysis~\cite{Lipstein:2019mpu}.
}
\begin{align}
	\LA J_2J_2J_2\RA|_{d=5} &= \frac{A_{\rm GR}}{K^3} \Big[2e_3^2+3(e_2+K^2)e_3K \\
	 &\qquad+3(e_2^2-3e_2K^2+K^4)K^2\Big]\,,\nn
	\LA J_2J_2J_2\RA|_{d=7} &=	\frac{3A_{\rm GR}}{K^4}\Big[2e_3^3+(4e_2-K^2)e_3^2K\\
&\qquad+5(e_2^2+3e_2K^2-3K^4)e_3K^2\nn
	&\qquad+5(e_2^3-6e_2^2K^2+5e_2K^4-K^6)K^3\Big]\, ,\nonumber
\end{align}
where $e_2\equiv k_1k_2+k_2k_3+k_3k_1$ and $e_3\equiv k_1k_2k_3$.

\subsection{Higher-Derivative Interactions}

For correlators arising from higher-derivative interactions, it turns out that it is most useful to consider the operator $\hF_{ab}$ to replace $\eps_a\cdot\k_b$ in \eqref{AF3}.
This is due to the property
\begin{align}
	\text{div}_a \hF_{a b}^\ell \LA\Phi\Phi\Phi\RA &\propto (d+2\ell-2-\Delta_\Phi)\times\cdots\, \label{divF}
\end{align}
after acting on a scalar correlator and taking the divergence, where we have just shown the proportionality constant. 
This property also holds for $\hF_{ab}^\ell \hF_{cd}^\ell \hF_{ef}^\ell$ with $a\ne c\ne e$, as long as the operators are grouped in this way.
This means that the resulting correlator becomes automatically divergenceless if we use the scalar seed with $\Delta_\Phi=d+2\ell-2$. 
The seed function choice then agrees with that in \eqref{JJJ1} and \eqref{JJJ2} due to the fact that both $\hF_{ab}^\ell \hF_{cd}^\ell \hF_{ef}^\ell$ and $\hn_\ell$ have an overall weight of $-\ell$, so that they give the correct weight for the conserved spin-$\ell$ current, $\Delta_{J_\ell} = d+\ell-2$.

The above discussion implies the following spin-$\ell$ formula for three-point functions from higher-derivative interactions:\footnote{A similar formula for the mixed $\ell$-$\ell$-0 correlator is
\begin{equation}
	\LA J_\ell J_\ell \Phi\RA  = \hF_{12}^\ell \hF_{23}^\ell \LA \Phi\Phi\Phi\RA_{\Delta_\Phi=d+2\ell-2}\,,
\end{equation}
which arises from the bulk coupling of the form $F^2\phi$ and its higher-spin generalizations.
}
\begin{equation}
	\hskip -7pt\LA J_\ell J_\ell J_\ell\RA_{\rm h.d.}\!= \hF_{12}^\ell \hF_{23}^\ell \hF_{31}^\ell \LA\Phi\Phi\Phi\RA_{\Delta_\Phi=d+2\ell-2}\hskip 1pt ,\label{hd}
\end{equation}
where we have picked a particular permutation of the operators.
For $\ell=1$, this agrees with \cite{Caron-Huot:2021kjy, Diwakar:2021juk}. 
For general spins, one should also check that \eqref{hd} comes purely from higher-derivative interactions. 
This is not immediately obvious in embedding space, since the divergenceless condition does not distinguish between the types of interactions. In momentum space, however, these higher-derivative contributions solve the homogeneous WT identity and are thus identically conserved~\cite{Baumann:2020dch,Baumann:2021fxj}, as well as having higher-order singularities in $K$ (see footnote~\ref{foot6}).
Different index choices of the operators in \eqref{hd}, e.g.,~$\hF_{12}\to\hF_{13}$, give the same correlator in embedding space but they generally differ in momentum space by local terms. 
We have explicitly checked that \eqref{hd} gives identically-conserved momentum-space correlators, up to local terms, for $\ell=2,3$.

\section{Conclusions}

What are the right kinematic variables for cosmological correlators? Given an amplitude in flat space, can we directly reconstruct the corresponding correlator in curved backgrounds?
In this letter, we have provided plausible answers to these questions for three-point functions in (A)dS$_{d+1}$. 
In particular, we used the weight-shifting operators developed in~\cite{Costa:2011dw, Karateev:2017jgd, Baumann:2019oyu} as basic kinematic building blocks to construct (A)dS three-point functions.
We introduced a normal ordering and the proper normalization of the weight-shifting operators, which allowed us to treat them as the (A)dS analogs of the kinematic variables of amplitudes.
Remarkably, the final differential representation of the gluon and graviton three-point functions in general dimensions has exactly the same kinematic structure as the corresponding amplitude, and thus exhibits a manifest double-copy relation. 

We have also seen the advantages of our hybrid embedding-momentum space approach. In embedding space, weight-shifting operators can be systematically constructed, and current conservation in general dimensions is simpler to prove. 
In momentum space, the amplitude-correlator connection is more direct, which allowed us to find a canonical normalization and ordering of the weight-shifting operators. 
Moreover, correlators of conserved currents have simple expressions in odd-$d$ momentum space, in which case the flat-space limit and the WT identity can serve as useful consistency conditions.
Our final results for (A)dS$_{d+1}$ three-point functions are valid in both embedding and momentum spaces.

The logical next step is to generalize our three-point double-copy construction to higher spins. 
In flat space, the spin-$\ell$ three-point amplitude is simply given by the $\ell$-th power of the spin-1 amplitude, so it is natural to expect a differential generalization of this in (A)dS.
Also, partially massless fields~\cite{Deser:1983mm, Deser:2001us,Dolan:2001ih,Deser:2003gw,Skvortsov:2006at,Joung:2012rv} are an intriguing class of particles unique to (A)dS with no flat-space analogs.
Correlators of partially massless fields are partially conserved on the boundary and have interesting features~\cite{Baumann:2017jvh,Goon:2018fyu,Sleight:2021iix}. 
It would be nice to understand what combinations of differential operators ensure partial current conservation. 

Our framework should provide a natural language for exploring the double copy of gluon and graviton correlators at higher multiplicities. 
This would involve enlarging the basis set of differential operators to include combinations of conformal generators, which are the (A)dS analogs of the Mandelstam variables~\cite{Roehrig:2020kck,Eberhardt:2020ewh, Herderschee:2022ntr, Diwakar:2021juk, Sivaramakrishnan:2021srm,  Cheung:2022pdk}. 
In addition, it would be interesting to work out a supersymmetric generalization and make contact with the existing double-copy formulation in AdS Mellin space~\cite{Zhou:2021gnu}.
A similar differential technique has proven useful in recent generalizations of the scattering equations~\cite{Cachazo:2013gna,Cachazo:2013hca,Cachazo:2013iea} to (A)dS~\cite{Roehrig:2020kck,Eberhardt:2020ewh, Gomez:2021qfd, Gomez:2021ujt, Armstrong:2022csc, Armstrong:2022vgl}, and it is worth exploring the synergy between related approaches.
Finally, our findings may also have implications to analytic studies of spinning correlators in conformal field theories with a weakly-coupled bulk dual~\cite{Meltzer:2017rtf,Karateev:2018oml,Caron-Huot:2021kjy}.

\vskip 5pt
\begin{acknowledgments}
{\it Acknowledgments.}\quad We thank 
Daniel Baumann and Savan Kharel for useful discussions, and Austin Joyce for comments on the draft. 
HL is supported by the Kavli Institute for Cosmological Physics at the University of Chicago through an endowment from the Kavli Foundation and its founder Fred Kavli.
\end{acknowledgments}

\bibliography{dc}

\end{document}